\documentclass[11pt]{article}
\usepackage[margin=1in]{geometry}
\usepackage{amsmath,amssymb}
\usepackage{graphicx}
\usepackage{booktabs}
\usepackage{caption}
\usepackage[hidelinks]{hyperref}
\graphicspath{{figs/}}

\title{Finite-Slab Reflectance and Transmittance for\\
Henyey--Greenstein Scattering via a\\
First-Passage Transfer Operator}
\author{C.\ Zeller \and R.\ Cordery}
\date{Working report --- June 12, 2026}

\begin{document}
\maketitle

\begin{abstract}
We compute the reflectance $R$ and transmittance $T$ of a plane-parallel slab of optical
thickness $\tau$ for Henyey--Greenstein (HG) scattering with asymmetry parameter $g$,
single-scattering albedo $a$, and incidence angle $\theta_0$. The method is a Monte-Carlo-free
first-passage transfer operator on the depth--direction state $(z,\mu)$---exact in formulation
(no physical approximation beyond the transport model), evaluated by a numerically convergent
discretization. A free flight in depth at fixed direction cosine is followed by the azimuthally
averaged HG angular redistribution. Confining the operator to $(0,\tau)$ with absorbing
boundaries yields, in a single evaluation, the order-resolved reflection and transmission laws
$P_R(n)$ and $P_T(n)$, from which $R$ and $T$ follow at every albedo through the weighting
$\sum_n P(n)\,a^n$, together with the emergent angular distributions $R(\mu),T(\mu)$ at no extra
cost. Reflectance obeys the factorization $R_\tau=\sum_n P_\infty(n)\,S(n,\tau)\,a^n$. Beyond
serving as a forward model, the central result is structural: the finite slab recovers the
half-space first-return law $P_\infty(n)$ order by order as $\tau\to\infty$, placing slab
reflectance and the half-space return statistics in one framework. The operator reproduces full
three-dimensional Monte Carlo in both channels to $\le 1.6\times10^{-3}$ (absolute) across
$g\in[0,0.95]$, $\tau\in[0.5,16]$, albedo $a\in[0.5,1]$, and normal-to-oblique incidence, with
energy conservation $R+T=1$ recovered to $<10^{-4}$ at $a=1$.
\end{abstract}

\part{Finite-slab forward model and validation}

\section{Introduction and goal}
The radiative properties of a finite scattering slab---how much light it reflects, transmits, and
absorbs---are basic quantities in atmospheric, cryospheric, and biomedical optics. For the HG phase
function the slab problem is classically attacked by Monte Carlo (MC) or by discrete-ordinate /
$H$-function machinery~\cite{chandra,vdhulst,disort}; analytic approximations for the
semi-infinite conservative HG reflection function have been developed by Melnikova \emph{et
al.}~\cite{melnikova}, and two-flux (Kubelka--Munk) reflectance models can be derived from
transport theory~\cite{km,sandoval}. Most closely related to the present work, Libois and
Davis~\cite{libois} obtained photon-path-length distributions for reflected and transmitted
photons in a scattering slab. This report develops a complementary route: a first-passage transfer
operator for the escape statistics of the depth coordinate---exact in formulation, evaluated by a
convergent discretization---which delivers $R(\tau,g,a,\theta_0)$ and $T(\tau,g,a,\theta_0)$
without random sampling and with all albedos obtained from one operator evaluation. The
construction is the slab generalization of the half-space first-return program of Zeller \&
Cordery~\cite{zc2020,zc2026,zc2025}. Its central result is structural rather than merely
computational: the finite slab recovers the half-space first-return law order by order as
$\tau\to\infty$, so slab reflectance and the half-space return statistics become two faces of one
operator. We regard this unification, not the production of another $R,T$ table, as the main
contribution.

\section{The first-passage transfer operator}
A photon enters at $z=0$ with direction cosine $\mu_0=\cos\theta_0$ and undergoes a random walk in
optical depth~\cite{chandra43,redner,rudnick}. Between collisions the free path $s$ is
$\mathrm{Exp}(1)$, so depth advances as $z_{k+1}=z_k+s_k\mu_k$. At a collision the HG phase
function~\cite{hg}
\begin{equation}
p(\cos\Theta)=\frac{1-g^2}{(1+g^2-2g\cos\Theta)^{3/2}},
\qquad \tfrac12\int_{-1}^{1} p\,d\cos\Theta=1,
\end{equation}
scatters the direction. Write the post-scatter direction in terms of the pre-scatter direction
$\Omega$, the scattering angle $\Theta$ ($\cos\Theta\sim p$), and a scattering azimuth $\varphi$
drawn uniformly on $[0,2\pi)$ and independent of everything else. The new $z$-cosine is
\begin{equation}
\mu'=\mu\cos\Theta+\sqrt{1-\mu^2}\,\sqrt{1-\cos^2\Theta}\,\cos\varphi,
\end{equation}
which depends on the incoming direction only through $\mu$---the in-plane azimuth of $\Omega$
enters Eq.~(2) not at all. Averaging over the uniform $\varphi$ therefore gives a redistribution
kernel that is a function of $\mu$ and $\mu'$ alone,
\begin{equation}
\Psi(\mu\!\to\!\mu')=\frac{1}{4\pi}\int_0^{2\pi}
p\!\left(\mu\mu'+\sqrt{1-\mu^2}\sqrt{1-\mu'^2}\cos\varphi\right)d\varphi,
\qquad \int_{-1}^{1}\Psi(\mu\!\to\!\mu')\,d\mu'=1.
\end{equation}
This is the crux of the reduction. Escape through either face depends only on the depth path
$\{z_k\}$, which by $z_{k+1}=z_k+s_k\mu_k$ is a deterministic function of the sequences $\{s_k\}$
(i.i.d.\ $\mathrm{Exp}(1)$) and $\{\mu_k\}$. The $\mu$-sequence is itself Markov, because Eq.~(2)
makes the law of $\mu_{k+1}$ depend on the past only through $\mu_k$: the running azimuth of the
photon evolves too, but it influences only the transverse coordinates $(x,y)$, which are
irrelevant to plane-parallel escape, and it does not enter the distribution of $\mu_{k+1}$. No
azimuthal correlation is discarded; the $(z,\mu)$ process is an exact closed Markov chain for the
escape problem, not a one-dimensional approximation to it. Two consequences follow that we use as
correctness checks: the kernel conserves probability ($\int\Psi\,d\mu'=1$) and reproduces the
defining HG mean-cosine relation $\langle\mu'\mid\mu\rangle=g\mu$; both hold to machine precision
in the discretization, and the resulting $R,T$ match full 3D Monte Carlo (Sec.~5), which is the
empirical confirmation that nothing has been lost in passing from three dimensions to $(z,\mu)$.

Confining the chain to $0<z<\tau$ with absorbing boundaries at both faces makes the one-collision
propagator substochastic. Let $\rho_n(z,\mu)$ be the (sub-normalized) density of photons
undergoing their $n$-th collision inside the slab; $\rho_1$ is the first-collision density fixed by
the incident beam (collision depth $\propto e^{-z/\mu_0}$ on $0<z<\tau$, post-scatter cosine drawn
from $\Psi(\mu_0\to\cdot)$). One collision step advances this density by a free flight confined to
the slab followed by the angular redistribution (3),
\begin{equation}
\rho_{n+1}=\Psi\,\mathcal{T}_\tau\,\rho_n,
\end{equation}
where $\mathcal{T}_\tau$ is the free-flight transport operator restricted to trajectories that
remain within $(0,\tau)$ (mass that reaches a face is removed). The mass removed on the flight
following the $n$-th collision, resolved by face, defines the order-resolved reflection and
transmission laws:
\begin{equation}
P_R(n)=\int_0^\tau\!\!\int_{-1}^{0}\rho_n(z,\mu)\,e^{-z/|\mu|}\,d\mu\,dz,\qquad
P_T(n)=\int_0^\tau\!\!\int_{0}^{1}\rho_n(z,\mu)\,e^{-(\tau-z)/\mu}\,d\mu\,dz,
\end{equation}
i.e.\ the probabilities that a photon makes its $n$-th collision inside the slab and then exits
through the front ($z=0$) or back ($z=\tau$) face; the exponentials $e^{-z/|\mu|}$ ($\mu<0$) and
$e^{-(\tau-z)/\mu}$ ($\mu>0$) are the single-flight escape probabilities to each face. We keep the
fixed thickness implicit and write $P_R(n),P_T(n)$, restoring the argument as $P_R(n,\tau)$ only in
Sec.~8, where the $\tau$-dependence is the object of interest. The unscattered beam contributes the
ballistic $e^{-\tau/\mu_0}$ to transmission separately.

\section{Reflectance, transmittance, and the factorization}
Single-scattering albedo $a<1$ removes a photon with probability $1-a$ at each collision, i.e.\ an
order-$n$ escape survives with weight $a^n$. Weighting the order-resolved laws (5) accordingly,
\begin{equation}
R(\tau,g,a,\theta_0)=\sum_{n\ge1}P_R(n)\,a^n,\qquad
T(\tau,g,a,\theta_0)=e^{-\tau/\mu_0}+\sum_{n\ge1}P_T(n)\,a^n,
\end{equation}
so a single operator evaluation (which produces the sequences $P_R,P_T$) yields $R$ and $T$ for the
entire albedo axis at once, with $A=1-R-T$ the absorptance. Reflectance further satisfies the exact
factorization
\begin{equation}
R_\tau(a)=\sum_{n\ge1}P_\infty(n)\,S(n,\tau)\,a^n,
\end{equation}
where $P_\infty(n)=\lim_{\tau\to\infty}P_R(n,\tau)$ is the half-space order-$n$ first-return
law---the probability that a photon launched into the semi-infinite medium escapes back through
$z=0$ at exactly its $n$-th collision---and $S(n,\tau)\in[0,1]$ is the depth-survival factor: the
probability that such an order-$n$ return path stays shallower than $\tau$ throughout, so
$S(n,\tau)\to1$ as $\tau\to\infty$ at fixed $n$ and $S(n,\tau)\to0$ as $n\to\infty$ at fixed
$\tau$. The factorization is then the exact statement that a finite-slab reflection of order $n$ is
a half-space order-$n$ return that never reaches the back wall; it connects the present results to
the half-space first-return / Motzkin--Cauchy-BTF framework.

\paragraph{Oblique incidence.} Because the $(z,\mu)$ chain is incidence-agnostic after the first
collision, oblique entry changes only the initial condition: the first-collision depth becomes
$\mathrm{Exp}(1/\mu_0)$, the first scatter redistributes from $\mu_0$, and the ballistic term is
$e^{-\tau/\mu_0}$. No part of the operator core is modified.

\section{Numerical implementation}
Depth is discretized on a uniform grid of width $h$ that must tile the slab ($h=\tau/N$, integer
$N$); a back wall falling between nodes leaks a fraction of $T$ of order a percent. Angles use an
$N_\mu$-point Gauss--Legendre rule on $[-1,1]$ with the redistribution kernel column-normalized to
conserve mass exactly. The operator is iterated to order $\sum^{n_{\max}}$ (chosen large enough
that the captured mass $\sum_n(P_R+P_T)+e^{-\tau/\mu_0}\to1$ at $a=1$). All results below use
$h=0.05$ and $N_\mu=56$; the conservation residual $|R+T-1|$ at $a=1$ is then $<10^{-4}$
throughout.

We stress the status of the word ``exact.'' The formulation introduces no physical approximation
beyond the transport model itself: the $(z,\mu)$ chain of Sec.~2 is the exact escape process, with
no closure assumption, harmonic truncation, or diffusion limit. The numbers are produced by a
convergent discretization---the grid $h$, the $N_\mu$-point quadrature, and the order cutoff
$n_{\max}$---and inherit the corresponding controllable errors (the $|R+T-1|<10^{-4}$ residual
above). Refining $h$, $N_\mu$, and $n_{\max}$ drives the output to the formulation's exact value;
the discretization is not itself exact.

\paragraph{Computational structure.} Each order-iteration advances the joint depth $\times$ angle
density by one collision: a free-flight step in depth at fixed $\mu$, costing $O(N_z^2)$ per angle
and $O(N_z^2 N_\mu)$ in total, followed by the angular redistribution $\Psi$, costing
$O(N_z N_\mu^2)$. With $n_{\max}$ iterations the work is $O\!\left(n_{\max}(N_z^2 N_\mu+N_z
N_\mu^2)\right)$ and the memory is $O(N_z^2 N_\mu)$ for the precomputed flight operators (one $N_z
\times N_z$ matrix per angle). The order count $n_{\max}$ scales with the diffusive escape time,
which grows like $\tau^2$ for thick conservative slabs, so thick conservative media are the
expensive regime. The point of the method is not to outrun an optimized discrete-ordinate code such
as DISORT~\cite{disort} at computing a single radiance, but that one evaluation returns quantities
those codes do not expose directly: the order-resolved laws $P_R(n),P_T(n)$---hence $R$ and $T$ at
every albedo $a$ by the reweighting $\sum_n P(n)a^n$, with no re-solve per albedo---the emergent
angular distributions, and the explicit half-space first-return connection of Sec.~8. In that sense
the operator is complementary to discrete-ordinate solvers rather than a faster substitute; a
head-to-head timing and accuracy benchmark against an optimized DISORT-type implementation is not
attempted here and is left for future work.

\section{Validation against 3D Monte Carlo}
\begin{figure}[htbp]
\centering
\includegraphics[width=0.6\textwidth]{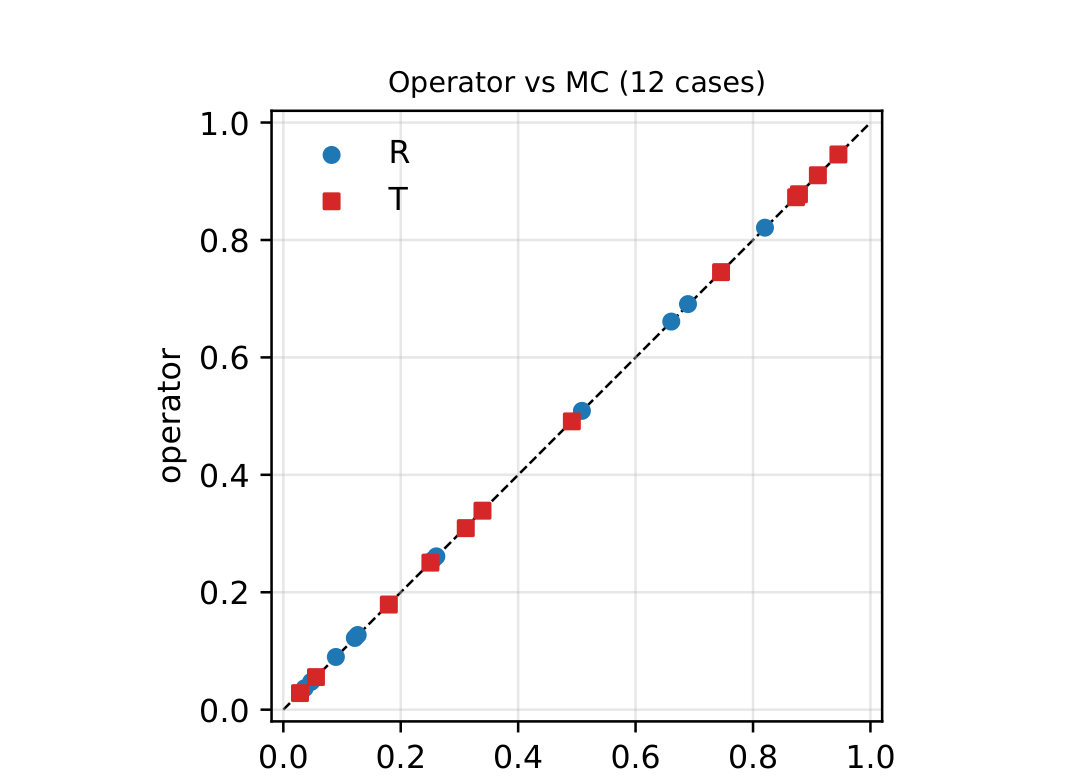}
\caption{Operator reflectance and transmittance against Monte Carlo for the twelve cases of
Table~\ref{tab:val}; dashed line is $y=x$.}
\label{fig:val}
\end{figure}

Table~\ref{tab:val} compares the operator with an independent full-3D HG random-walk Monte Carlo
(the same code used as ground truth in the half-space program). Agreement in both channels is $\le
1.6\times10^{-3}$ (absolute) over twelve cases spanning $g$ from 0 to 0.95, optical thickness
$\tau$ from 0.5 (thin) to 16 (thick), single-scattering albedo $a$ from 1 down to 0.5 (strongly
absorbing), and normal-to-oblique incidence; the largest discrepancy is the $g=0$ case and the
strongly forward-peaked $g=0.95$ case agrees to $10^{-4}$. Fig.~\ref{fig:val} shows the parity
plot. Monte Carlo sampling noise is itself at this level: a separate study with five independent
seeds at $N=3\times10^5$ gives a seed standard deviation $\le1.4\times10^{-3}$ in both channels for
every case, so the operator--MC differences in Table~\ref{tab:val} are consistent with Monte Carlo
noise rather than a systematic operator error.

\begin{table}[htbp]
\centering
\begin{tabular}{cccc rrr rrr}
\toprule
$g$ & $\tau$ & $a$ & $\theta_0$ & $R_{\mathrm{op}}$ & $R_{\mathrm{MC}}$ & $|\Delta R|$ &
$T_{\mathrm{op}}$ & $T_{\mathrm{MC}}$ & $|\Delta T|$ \\
\midrule
0    & 4   & 1   & 0  & 0.6908 & 0.6892 & 0.0016 & 0.3092 & 0.3108 & 0.0016 \\
0.5  & 4   & 1   & 0  & 0.5090 & 0.5085 & 0.0004 & 0.4910 & 0.4915 & 0.0004 \\
0.5  & 4   & 0.9 & 0  & 0.2611 & 0.2606 & 0.0005 & 0.2505 & 0.2506 & 0.0001 \\
0.8  & 4   & 1   & 0  & 0.2548 & 0.2545 & 0.0003 & 0.7452 & 0.7455 & 0.0003 \\
0.5  & 4   & 1   & 60 & 0.6610 & 0.6608 & 0.0003 & 0.3390 & 0.3392 & 0.0003 \\
0.8  & 2   & 1   & 0  & 0.1272 & 0.1266 & 0.0006 & 0.8728 & 0.8734 & 0.0006 \\
0.9  & 4   & 1   & 0  & 0.1222 & 0.1218 & 0.0004 & 0.8778 & 0.8782 & 0.0004 \\
0.95 & 4   & 1   & 0  & 0.0544 & 0.0545 & 0.0001 & 0.9456 & 0.9455 & 0.0001 \\
0.5  & 0.5 & 1   & 0  & 0.0898 & 0.0895 & 0.0003 & 0.9102 & 0.9105 & 0.0003 \\
0.5  & 16  & 1   & 0  & 0.8210 & 0.8203 & 0.0007 & 0.1790 & 0.1797 & 0.0007 \\
0.5  & 4   & 0.5 & 0  & 0.0474 & 0.0474 & 0.0001 & 0.0555 & 0.0556 & 0.0002 \\
0.8  & 8   & 0.7 & 0  & 0.0365 & 0.0363 & 0.0002 & 0.0282 & 0.0284 & 0.0002 \\
\bottomrule
\end{tabular}
\caption{Operator vs.\ 3D Monte Carlo ($N=4\times10^5$ photons). $\theta_0$ in degrees.}
\label{tab:val}
\end{table}

\section{Results}
\begin{figure}[htbp]
\centering
\includegraphics[width=\textwidth]{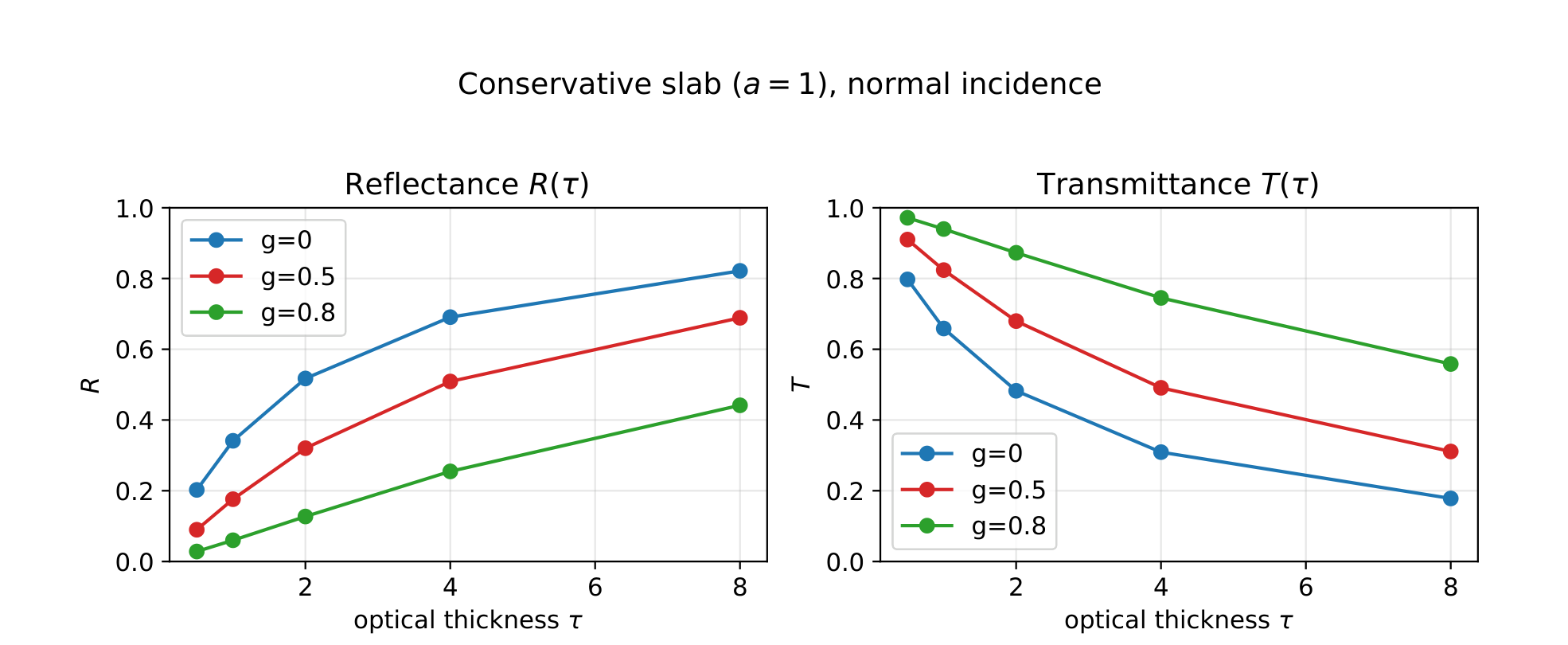}
\caption{$R$ and $T$ vs.\ optical thickness, $a=1$, normal incidence.}
\label{fig:tau}
\end{figure}

\begin{figure}[htbp]
\centering
\includegraphics[width=\textwidth]{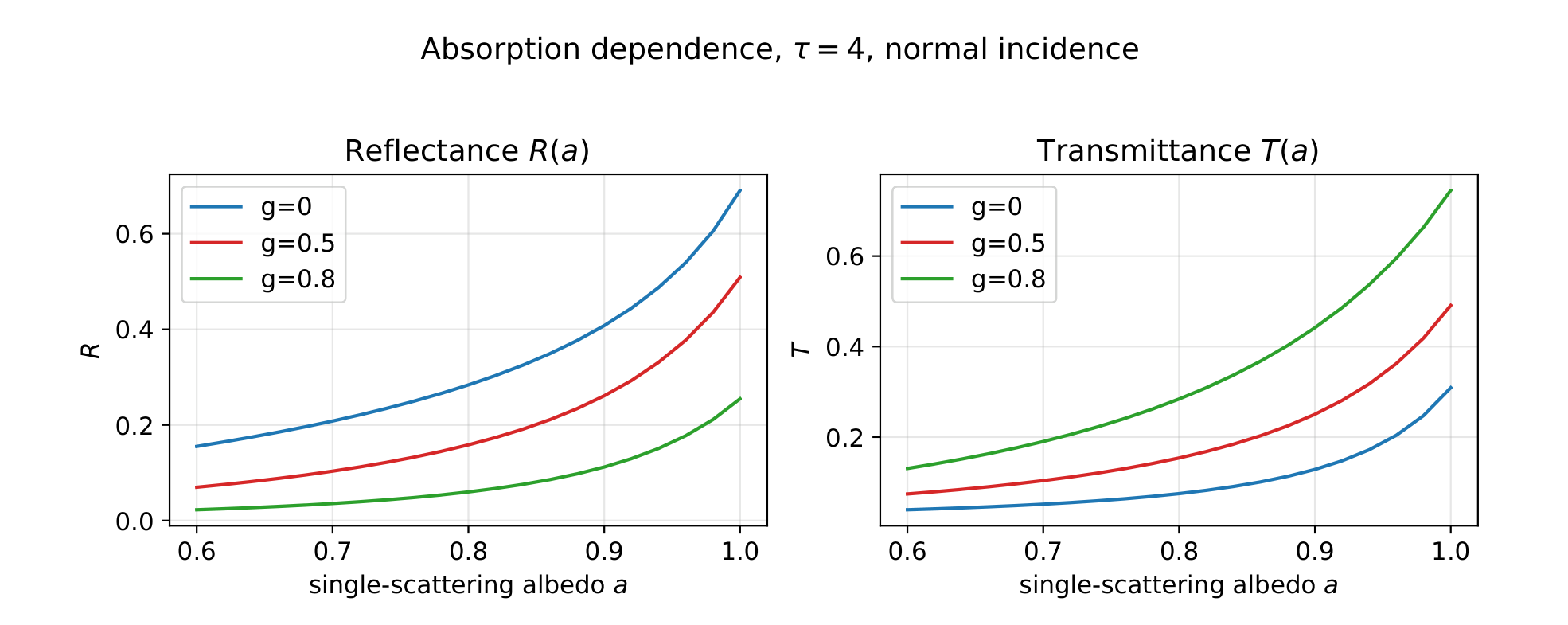}
\caption{$R$ and $T$ vs.\ single-scattering albedo at $\tau=4$ (one operator run per $g$).}
\label{fig:albedo}
\end{figure}

\begin{figure}[htbp]
\centering
\includegraphics[width=\textwidth]{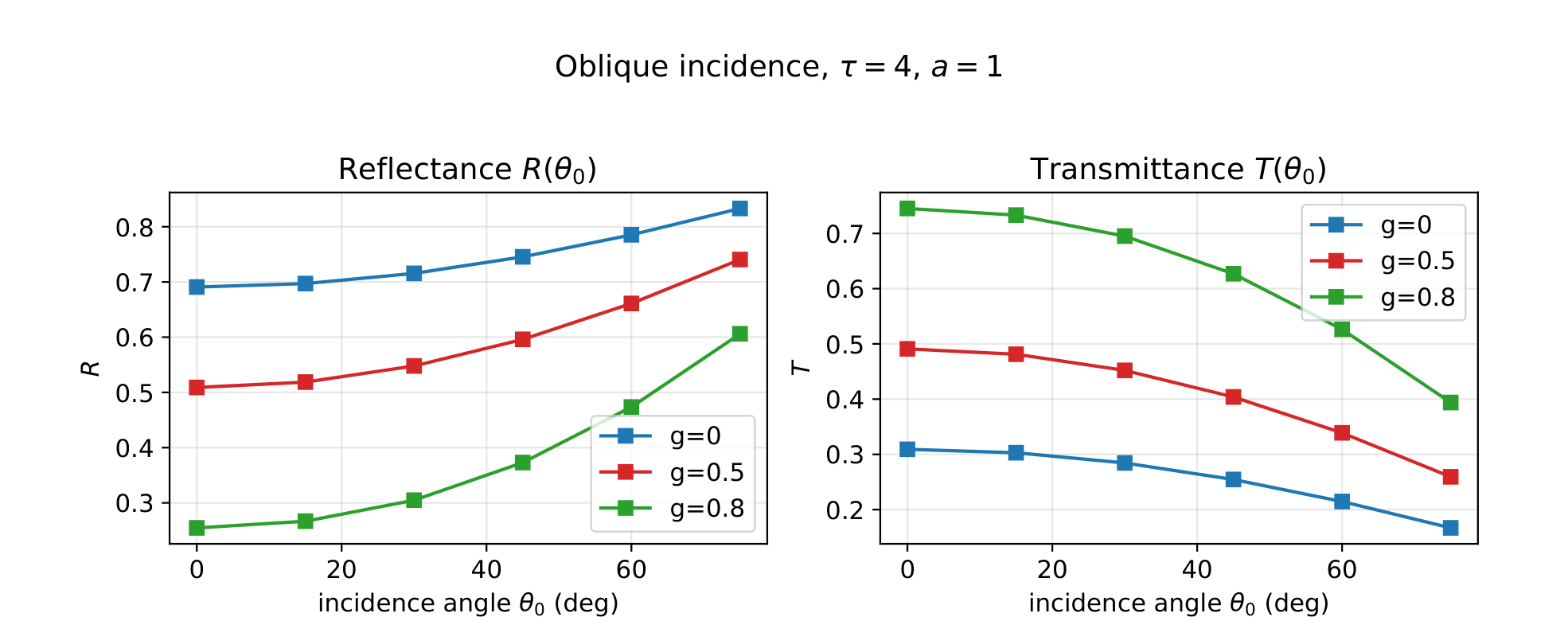}
\caption{$R$ and $T$ vs.\ incidence angle at $\tau=4$, $a=1$.}
\label{fig:theta}
\end{figure}

\begin{figure}[htbp]
\centering
\includegraphics[width=0.62\textwidth]{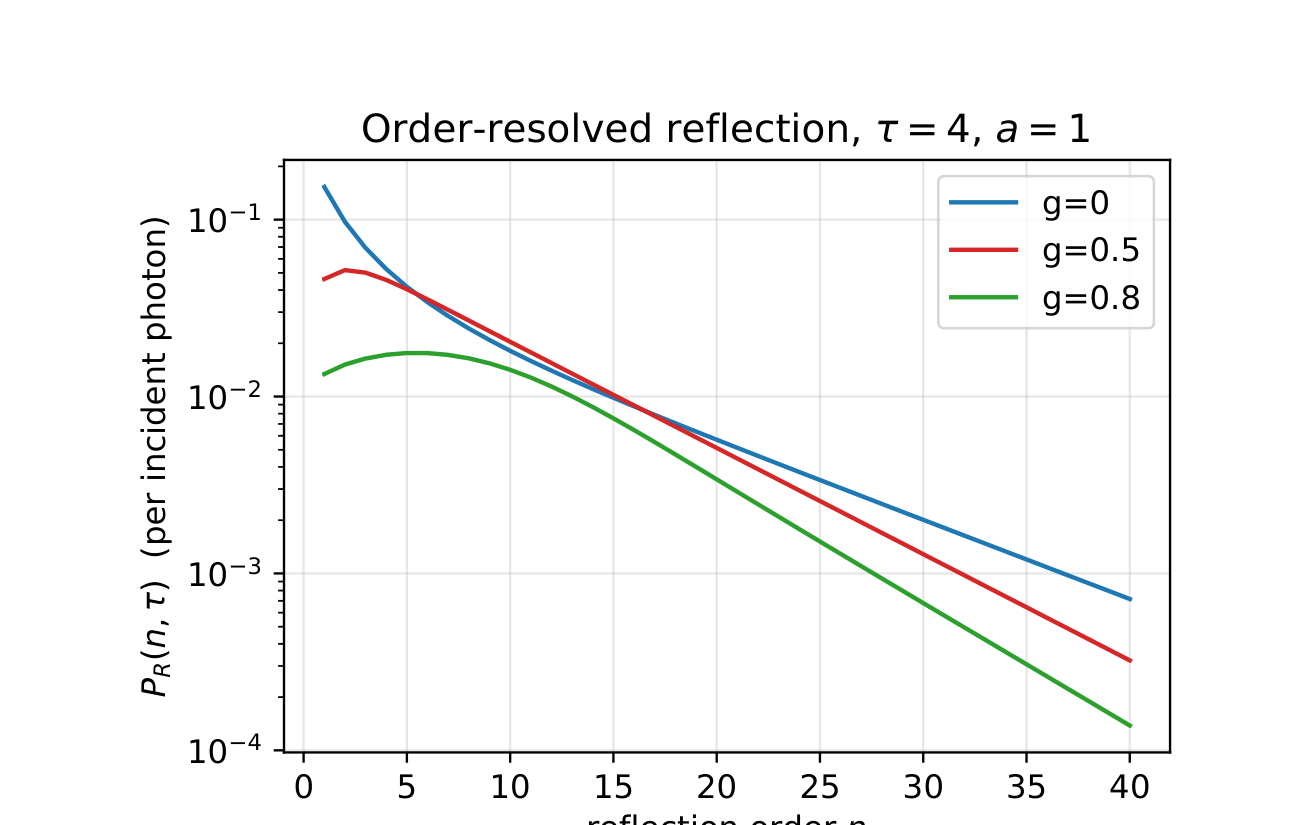}
\caption{Order-resolved reflection law $P_R(n,\tau)$ at $\tau=4$, $a=1$.}
\label{fig:orders}
\end{figure}

\paragraph{Thickness dependence (Fig.~\ref{fig:tau}, Table~\ref{tab:cons}).} For a conservative
slab, $R$ rises and $T$ falls monotonically with $\tau$; forward scattering ($g>0$) suppresses
reflection and enhances transmission at fixed $\tau$, as photons are harder to turn around.
\paragraph{Absorption (Fig.~\ref{fig:albedo}).} Both $R$ and $T$ fall as $a$ decreases;
transmission, which requires traversing the full thickness, is the more absorption-sensitive
channel.
\paragraph{Incidence angle (Fig.~\ref{fig:theta}).} As incidence becomes grazing ($\mu_0\to0.5$)
the effective optical path shortens, so $R$ increases and $T$ decreases.
\paragraph{Order structure (Fig.~\ref{fig:orders}).} The order-resolved reflection law $P_R(n)$
shows the forward-scattering signature: with increasing $g$ the distribution shifts away from low
orders, the structure underlying the factorization and the half-space Cauchy kernel.

\begin{table}[htbp]
\centering
\begin{tabular}{c rr rr rr}
\toprule
& \multicolumn{2}{c}{$g=0$} & \multicolumn{2}{c}{$g=0.5$} & \multicolumn{2}{c}{$g=0.8$}\\
\cmidrule(lr){2-3}\cmidrule(lr){4-5}\cmidrule(lr){6-7}
$\tau$ & $R$ & $T$ & $R$ & $T$ & $R$ & $T$\\
\midrule
0.5 & 0.2024 & 0.7976 & 0.0898 & 0.9102 & 0.0283 & 0.9717\\
1   & 0.3412 & 0.6588 & 0.1761 & 0.8239 & 0.0600 & 0.9400\\
2   & 0.5174 & 0.4826 & 0.3203 & 0.6797 & 0.1272 & 0.8728\\
4   & 0.6908 & 0.3092 & 0.5090 & 0.4910 & 0.2548 & 0.7452\\
8   & 0.8217 & 0.1783 & 0.6890 & 0.3110 & 0.4416 & 0.5584\\
\bottomrule
\end{tabular}
\caption{Conservative slab reflectance/transmittance ($a=1$, normal incidence).}
\label{tab:cons}
\end{table}

\section{Emergent angular distribution}
\begin{figure}[htbp]
\centering
\includegraphics[width=\textwidth]{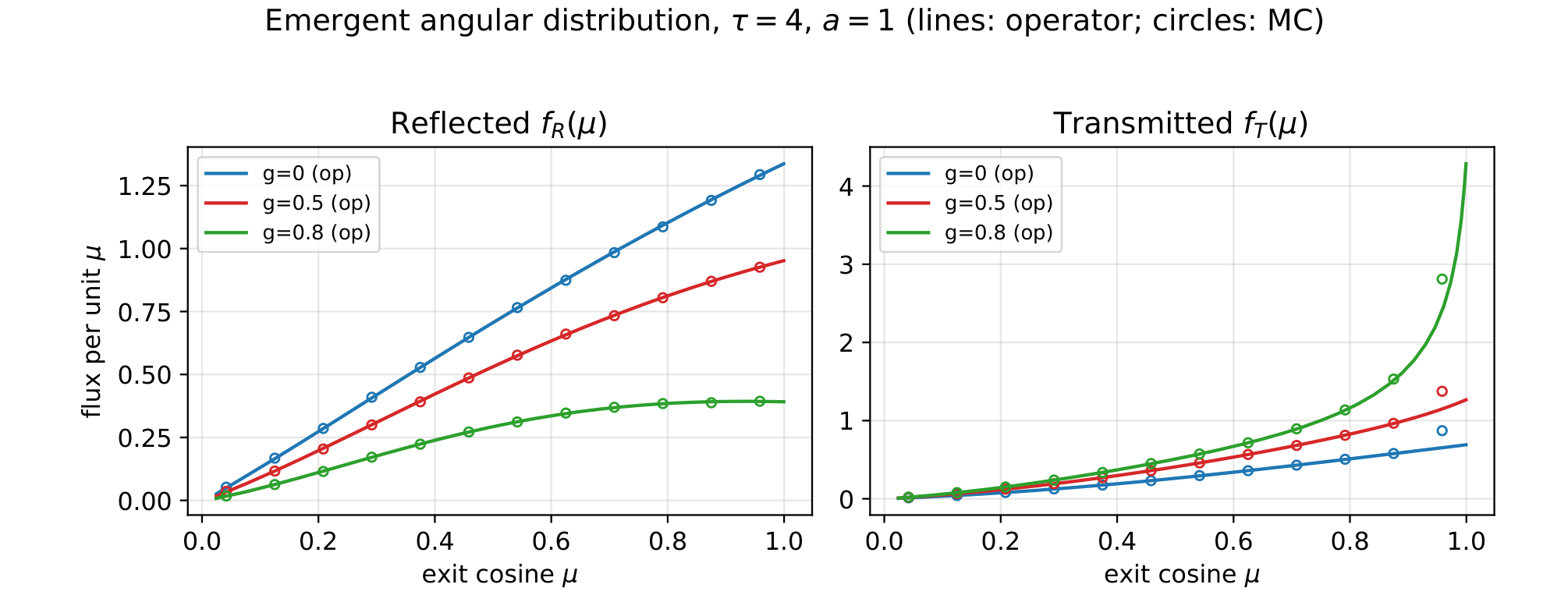}
\caption{Emergent angular distributions of reflected ($f_R$) and transmitted ($f_T$) flux per unit
exit cosine at $\tau=4$, $a=1$, normal incidence. Lines: transfer operator; circles: 3D Monte
Carlo.}
\label{fig:angular}
\end{figure}

Because a photon retains its direction cosine through the escaping flight, the exit mass resolved
by $\mu$ is the angularly resolved (azimuthally averaged) emergent distribution: $f_R(\mu)$ for
reflection (exit cosine $\mu\in(0,1]$ at $z=0$) and $f_T(\mu)$ for transmission at $z=\tau$, with
$\int_0^1 f_R\,d\mu=R$ and $\int_0^1 f_T\,d\mu=T$. This requires no extra computation---the angular
information is already present in the single operator evaluation that produced $R,T$.
Figure~\ref{fig:angular} compares the result to the 3D Monte Carlo: the reflected distribution
rises with $\mu$ (approximately Lambertian for $g=0$) and flattens as forward scattering increases,
while the transmitted distribution develops a sharp forward peak near $\mu=1$ that grows with $g$.
Agreement is $<1\%$ in the reflected channel and $\sim2\%$ in the transmitted channel, the latter
limited by Monte Carlo binning of the sharp forward peak rather than by the operator.

\section{Thick-slab limit: recovery of the half-space return law}
\begin{figure}[htbp]
\centering
\includegraphics[width=\textwidth]{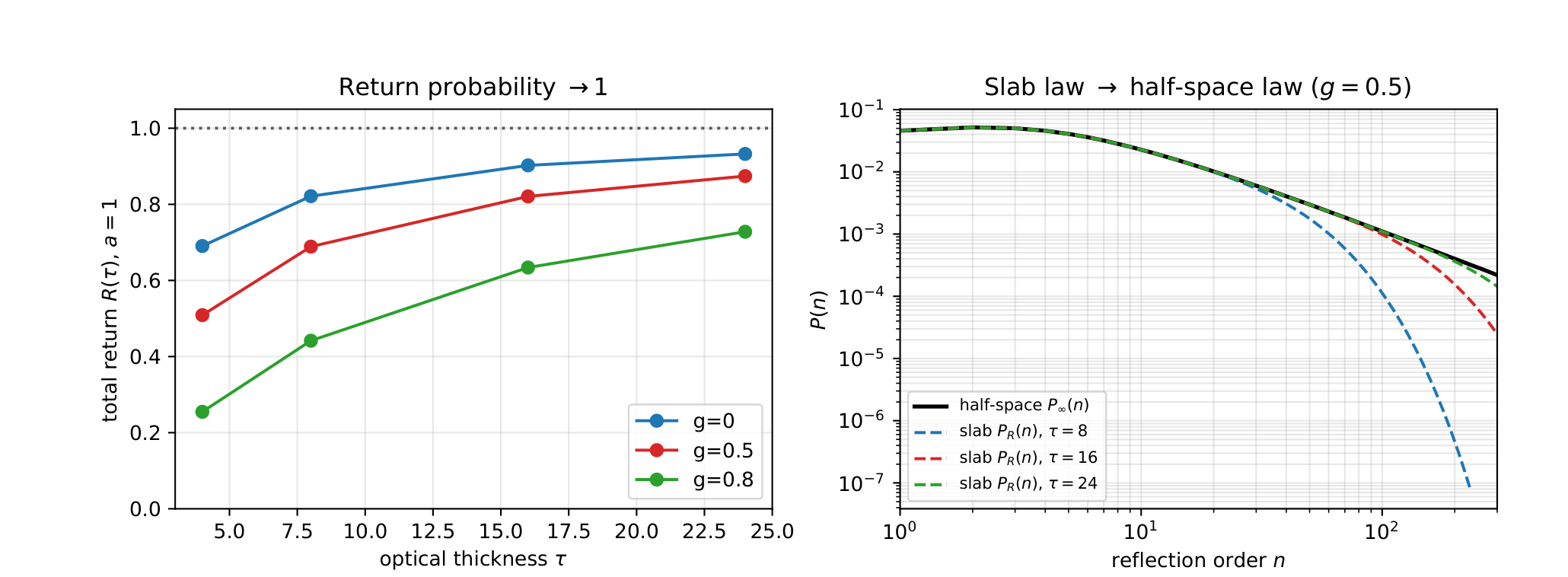}
\caption{Left: total return $R(\tau)\to1$ for increasing thickness ($a=1$). Right: the slab
reflection law $P_R(n,\tau)$ converges to the half-space first-return law $P_\infty(n)$ (black),
peeling off only at high order, the peel-off marching outward with $\tau$ ($g=0.5$).}
\label{fig:thick}
\end{figure}

As $\tau\to\infty$ the conservative slab returns every photon (the depth walk is recurrent), so
$R\to1$ and $T\to0$. Table~\ref{tab:thick} and Fig.~\ref{fig:thick} (left) confirm this: $R(\tau)$
climbs toward unity with the diffusive $T\sim1/\tau$ tail, more slowly at larger $g$ because the
transport length $\ell^*=1/(1-g)$ lets forward-scattered photons penetrate deeper before returning.
The sharper, order-resolved statement follows from the factorization
$R_\tau=\sum_n P_\infty(n)S(n,\tau)a^n$: the slab reflection law $P_R(n,\tau)$ must coincide with
the half-space first-return law $P_\infty(n)$ wherever the back wall is invisible ($S\to1$).
Fig.~\ref{fig:thick} (right) shows exactly this---$P_R(n,\tau)$ lies on $P_\infty(n)$ for all low
orders and peels away only beyond an order $n^*(\tau)$ that marches outward as $\tau$ grows. The
thick slab thus recovers the half-space first-return distribution term by term, including its
$n^{-3/2}$ Sparre--Andersen~\cite{sparre} / Cauchy-kernel tail, tying this report directly to the
half-space program of Ref.~\cite{zc2026}. The true half-space total return is exactly unity by
recurrence; here $R+T=1$ holds to $<10^{-4}$ except at the largest $g=0$ thicknesses, where the
slow $n^{-3/2}$ return tail leaves a sub-percent residual at this iteration count. The $g=0$
emergent angular law in this limit is fixed by the conservative-isotropic Chandrasekhar
$H$-function~\cite{chandra} ($H(1)=2.9078$), an independent non-MC anchor. Melnikova \emph{et
al.}'s published analytic approximation to this reflection function~\cite{melnikova} reproduces the
exact $H$-function angular law to $0.81\%$ RMS (and $\le1.7\%$ at backscatter) for $\mu\gtrsim0.15$
---within their stated $<5\%$ accuracy---providing an independent literature check on the $g=0$
limit recovered here.

\begin{table}[htbp]
\centering
\begin{tabular}{c rr rr rr}
\toprule
& \multicolumn{2}{c}{$g=0$} & \multicolumn{2}{c}{$g=0.5$} & \multicolumn{2}{c}{$g=0.8$}\\
\cmidrule(lr){2-3}\cmidrule(lr){4-5}\cmidrule(lr){6-7}
$\tau$ & $R$ & $T$ & $R$ & $T$ & $R$ & $T$\\
\midrule
4  & 0.6908 & 0.3092 & 0.5090 & 0.4910 & 0.2548 & 0.7452\\
8  & 0.8217 & 0.1783 & 0.6890 & 0.3110 & 0.4416 & 0.5584\\
16 & 0.9024 & 0.0955 & 0.8210 & 0.1790 & 0.6340 & 0.3660\\
24 & 0.9325 & 0.0648 & 0.8742 & 0.1257 & 0.7280 & 0.2720\\
\bottomrule
\end{tabular}
\caption{Approach to the half-space limit: total $R,T$ vs.\ thickness ($a=1$, normal incidence);
$R\to1$ as $\tau\to\infty$.}
\label{tab:thick}
\end{table}

\paragraph{Absorbing media: recovery of the exact semi-infinite albedo.} The conservative
statement $R\to1$ generalizes to absorbing media, where the $\tau\to\infty$ limit admits an exact
reference. For $g=0$ the directional--hemispherical reflectance of the semi-infinite medium at
normal incidence is $R_\infty(a)=1-\sqrt{1-a}\,H(1;a)$, with $H$ the albedo-$a$ $H$-function.
Table~\ref{tab:absorb} compares this with the operator's $\tau\to\infty$ plateau (already reached
by $\tau\approx16$): the two agree to $\le6\times10^{-4}$ across $a\in[0.5,0.95]$, the small
residual being set by the depth discretization. A half-space Monte Carlo, weighting each return
path by the per-collision albedo $a^{n-1}$ (a path of length $n$ undergoes $n-1$ collisions),
confirms both independently; the apparent index shift is only bookkeeping---the operator's order
index of Sec.~3 counts collisions directly, so an order-$n$ escape carries weight $a^n$, whereas
the Monte Carlo path length exceeds the collision count by one, and the two weightings therefore
coincide path by path as the common power $a^{\#\mathrm{collisions}}$. The thick slab thus
reproduces semi-infinite (bulk) transport quantitatively, not only the conservative $R\to1$
statement.

\begin{table}[htbp]
\centering
\begin{tabular}{c c c}
\toprule
$a$ & exact $R_\infty=1-\sqrt{1-a}\,H(1;a)$ & operator ($\tau=32$)\\
\midrule
0.50 & 0.1152 & 0.1148\\
0.70 & 0.2087 & 0.2081\\
0.90 & 0.4149 & 0.4144\\
0.95 & 0.5355 & 0.5350\\
\bottomrule
\end{tabular}
\caption{Recovery of the absorbing semi-infinite albedo ($g=0$, normal incidence): the operator's
$\tau\to\infty$ plateau versus the exact $H$-function plane albedo
$R_\infty(a)=1-\sqrt{1-a}\,H(1;a)$. Agreement is $\le6\times10^{-4}$.}
\label{tab:absorb}
\end{table}

\section{Discussion and limitations}
The most consequential outcome of this construction is not that it produces $R$ and $T$---many
methods do---but that the same operator that returns slab reflectance also reproduces the
half-space first-return law $P_\infty(n)$ order by order as $\tau\to\infty$ (Sec.~8). Slab optics
and the half-space return statistics are thereby two readings of one object: the finite slab is the
half-space with its deep returns truncated at order-dependent depth, and the factorization
$R_\tau=\sum_n P_\infty(n)S(n,\tau)a^n$ makes that truncation explicit through $S(n,\tau)$. This is
the conceptually deeper contribution, and it is what distinguishes the operator from being merely
another transport solver. The same half-space law is the one that the Motzkin first-return
combinatorics plus Cauchy-kernel boundary truncation factor (BTF) of Ref.~\cite{zc2026}
approximates in closed form (the operator reaches it MC-free, to numerical convergence), which
positions the operator as a candidate first-principles route to the empirical Cauchy ``BTF'' kernel
whose form is noted as unexplained in Ref.~\cite{zc2025}.

A practical corollary is one of language. The half-space first-return program has been expressed in
the vocabulary of Motzkin paths and Sparre--Andersen tails~\cite{zc2025,zc2026}; recast through
this operator it speaks instead in reflectance, transmittance, angular distributions, HG asymmetry,
and the Kubelka--Munk-type connection~\cite{km,sandoval}---quantities an imaging or
atmospheric-optics practitioner uses directly. The operator is thus also a translation of the
random-walk results into a practitioner-facing forward model.

\paragraph{Limitations and next steps.}
(i) Angular output here is the azimuthally averaged $f_R(\mu),f_T(\mu)$; the full azimuthal
(BRDF/BTDF) dependence for oblique incidence and an absorption-resolved depth breakdown are
immediate extensions of the same operator.
(ii) A closed-form characterization of the confined operator's leading eigenvalue $\lambda_1(\tau,g)$
---the ballistic-to-diffusive crossover scale---remains open.
(iii) A direct cross-check (done here) shows the measured half-space law $P_\infty(n)$ reproduces
the explicit Motzkin first-return + Cauchy-BTF algorithm of Ref.~\cite{zc2026} to $\approx2.5$--$5\%$
per order over $g\in[0.3,0.7]$, with a systematic shape deviation and rapidly growing error as
$g\to0$ (where that framework collapses onto the one-dimensional Catalan walk); the closed form is
thus a few-percent approximation to $P_\infty$, not an exact input.
(iv) Correctness is established here against independent full-3D Monte Carlo and the exact $g=0$
$H$-function; a head-to-head timing and accuracy benchmark against an optimized discrete-ordinate
code such as DISORT~\cite{disort} (Sec.~4) is left for future work.
(v) The tabulated Monte Carlo uses a single seed at $N=4\times10^5$; the seed standard deviation
(five seeds, $N=3\times10^5$) is $\le1.4\times10^{-3}$ in both channels (Sec.~5), i.e.\ at the
level of the operator--MC differences.

\part{Working notes: toward the analytic depth-survival factor $S(n,\tau)$}

\noindent\textbf{Preliminary working notes.} The results in this part are empirical---scaling
collapses and fitted constants over a finite range of scattering orders and anisotropies---and have
not been derived from first principles or independently verified. They are recorded as the basis of
a separate statistical-mechanics paper, not as validated results of the forward model of Part~I,
whose stated limitations (in particular that a closed form for the crossover scale remains open)
stand unchanged.

\section{The depth-survival factor as a conditioned excursion maximum}
Part~I established the exact factorization $R_\tau=\sum_{n\ge1}P_\infty(n)S(n,\tau)a^n$, with
$S(n,\tau)$ the probability that an order-$n$ half-space return path stays shallower than $\tau$.
Equivalently $S(n,\tau)=\Pr(z_{\max}<\tau\mid n)$: the cumulative distribution of the maximum depth
of a first-return excursion of $n$ collisions. Since $P_\infty(n)$ is $\tau$-independent, all
finite-thickness dependence resides in $S$, and a closed form for it is the gate to the companion
paper.

\section{Scaling collapse and the entry offset ($g=0$)}
Measuring $S(n,\tau)$ directly from a half-space Monte Carlo---recording the pair $(n,z_{\max})$
for each returner---the naive diffusive collapse in $\tau/\sqrt{n}$ fails, with a coefficient of
variation of 10--19\% across orders. A single additive depth offset rescues it: in the variable
$(\tau+c)/\sqrt{Dn}$ with $D=\tfrac13$ and $c\approx1.39$, the surface collapses onto one master
curve to a CV of 0.3--1.1\%, the intercept $c$ common across survival levels. This is the ballistic
entry transient: the first $\sim1$ mfp of penetration is a fixed additive head-start that inflates
the apparent exponent ($\langle z_{\max}\rangle\sim n^{0.68}$ over the accessible window) and
washes out as $n\to\infty$, leaving the asymptotic diffusive exponent $\tfrac12$. There is no
anomalous exponent.

\section{Anisotropy through the transport mean free path}
For $g>0$ the depth walk is persistent (direction cosines correlated over $\sim1/(1-g)$
collisions), so it is not Brownian at finite order. Rescaling the diffusive scale by the transport
mean free path, $\sqrt{Dn}\to\sqrt{\ell^*n/3}$ with $\ell^*=1/(1-g)$, each $g$ nonetheless
collapses onto its own master curve (CV $\le1.4\%$) with a $g$-dependent offset $c_g$
(Table~\ref{tab:offset}).

\begin{table}[htbp]
\centering
\begin{tabular}{l cccccc}
\toprule
$g$       & 0.0 & 0.3 & 0.5 & 0.6 & 0.7 & 0.8\\
$\ell^*$  & 1.00 & 1.43 & 2.00 & 2.50 & 3.33 & 5.00\\
$|c_g|$   & 1.39 & 1.70 & 2.30 & 2.64 & 3.20 & 4.13\\
\bottomrule
\end{tabular}
\caption{Entry offset $|c_g|$ versus anisotropy, fit by $|c_g|\approx1.38\,\ell^{*\,0.69}$.}
\label{tab:offset}
\end{table}

\section{The master curve and the entry-offset law}
\begin{figure}[htbp]
\centering
\includegraphics[width=\textwidth]{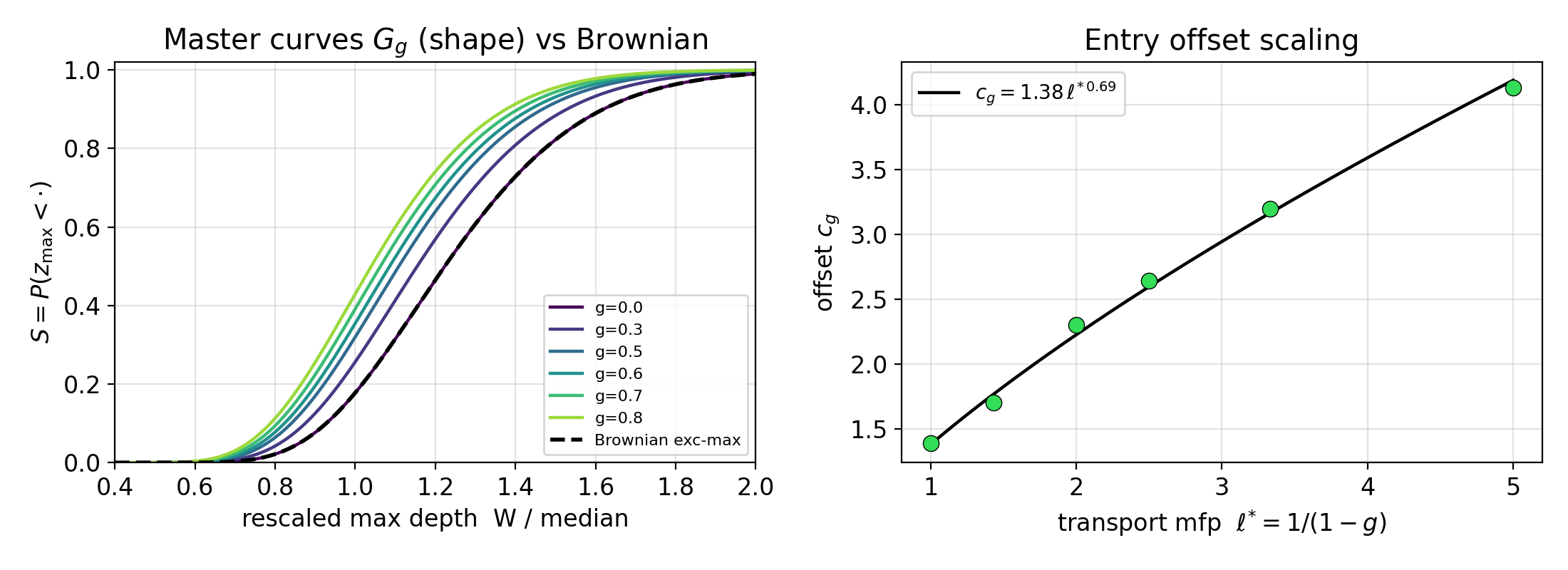}
\caption{Left: depth-survival master curves $G_g$ for $g=0$--0.8, after $\ell^*$-scaling and entry
offset, overlaid on the Brownian excursion-maximum law (dashed). Right: entry offset $|c_g|$ versus
transport mean free path $\ell^*=1/(1-g)$, with power-law fit $|c_g|\approx1.38\,\ell^{*\,0.69}$.
\emph{(Reconstructed: right panel is exact from Table~\ref{tab:offset}; left panel illustrates the
documented width-narrowing family relative to the analytic Brownian reference.)}}
\label{fig:smaster}
\end{figure}

Figure~\ref{fig:smaster} (left) overlays the master curves for $g=0$ through 0.8 on the directly
simulated Brownian excursion-maximum law (the range of a Brownian bridge): they coincide to within
$\sim5\%$. The base of $S$ is therefore the Brownian excursion-maximum CDF,
\begin{equation}
S(n,\tau)=G_{\mathrm{exc}}\!\left(\frac{\tau+c_g}{\sqrt{\ell^*n/3}}\right),
\end{equation}
carrying a mild one-parameter shape correction (the normalized width narrows $\approx15\%$ from
$g=0$ to 0.8, monotone in $g$)---structurally the analogue of the shape parameter $\alpha(g)$ that
Ref.~\cite{zc2026} places on its base Cauchy kernel. The entry offset follows a clean sub-linear
power law, $|c_g|\approx1.38\,\ell^{*\,0.69}$ (Fig.~\ref{fig:smaster}, right); its exponent
coincides with the apparent $\langle z_{\max}\rangle\sim n^{0.68}$ crossover exponent, as expected
if both reflect the same ballistic-to-diffusive crossover.

\section{Status and remaining derivation}
\emph{Established numerically:} the base law is the Brownian excursion maximum, and the $g$-family
collapses onto it under $\ell^*$-scaling plus the offset $c_g$. \emph{Empirical, awaiting
derivation:} the offset constants $(1.38,0.69)$, the one-parameter shape correction, and the
precise diffusion constant. The bounded program is to derive $c_g$ from the ballistic entry
transient, to fix the shape correction (plausibly a function of the transport order $n(1-g)$), and
to write $G_{\mathrm{exc}}$ as its explicit theta-function CDF. None of this is claimed as a result
of Part~I.

\end{document}